\documentclass[superscriptaddress,aps,pra,reprint,longbibliography]{revtex4-1}
\usepackage{etex}
\usepackage{graphicx}
\usepackage{caption}
\usepackage{subcaption}
\usepackage{mathrsfs}
\usepackage{amsfonts}
\usepackage{dsfont}
\usepackage{times}
\usepackage{amsmath}
\usepackage{amsthm}
\usepackage{leftidx}
\usepackage{tikz}
\usepackage{tikz-network}
\usepackage{color}
\usepackage[ bookmarks=true, colorlinks, linkcolor=blue, urlcolor=blue, citecolor=blue, plainpages=false, pdfpagelabels, final, breaklinks=true ]{hyperref}

\usepackage{mathtools}
\usepackage{array}
\usepackage{multirow}
\usepackage{bbold}
\usepackage{ragged2e}%

\newcommand{\ket}[1]{|#1\rangle}
\newcommand{\bra}[1]{\langle#1|}

\newcommand*{\tn}[1]{{\textnormal{#1}}}

\newcolumntype{L}[1]{>{\raggedright\let\newline\\\arraybackslash\hspace{0pt}}m{#1}}
\newcolumntype{C}[1]{>{\centering\let\newline\\\arraybackslash\hspace{0pt}}m{#1}}
\newcolumntype{R}[1]{>{\raggedleft\let\newline\\\arraybackslash\hspace{0pt}}m{#1}}

\theoremstyle{definition}

\DeclareRobustCommand{\orcidicon}{%
	\begin{tikzpicture}
	\draw[lime, fill=lime] (0,0) 
	circle [radius=0.16] 
	node[white] {{\fontfamily{qag}\selectfont \tiny ID}};
	\draw[white, fill=white] (-0.0625,0.095) 
	circle [radius=0.007];
	\end{tikzpicture}
	\hspace{-2mm}
}
\foreach \x in {A, ..., Z}{%
	\expandafter\xdef\csname orcid\x\endcsname{\noexpand\href{https://orcid.org/\csname orcidauthor\x\endcsname}{\noexpand\orcidicon}}
}

\begin{document}
\title{Approximating Maximum Independent Set on Rydberg Atom Arrays Using Local Detunings}
\author{Hyeonjun Yeo\orcidA{}}
\email{duguswns11@snu.ac.kr}
\affiliation{Department of Physics and Astronomy, Seoul National University, Seoul 08826, Korea}
\author{Ha Eum Kim\orcidB{}}
\email{hekim007@korea.ac.kr}
\affiliation{Department of Physics, Korea University, Seoul 02841, Korea}
\author{Kabgyun Jeong\orcidC{}}
\email{kgjeong6@snu.ac.kr}
\affiliation{Research Institute of Mathematics, Seoul National University, Seoul 08826, Korea}
\affiliation{School of Computational Sciences, Korea Institute for Advanced Study, Seoul 02455, Korea}

\date{\today}

\begin{abstract}
Rydberg atom arrays operated by a quantum adiabatic principle are among the most promising quantum simulating platforms due to their scalability and long coherence time. From the perspective of combinatorial optimization, they offer an efficient solution for an intrinsic maximum independent set problem because of the resemblance between the Rydberg Hamiltonian and the cost function of the maximum independent set problem. In this study, a strategy is suggested to approximate maximum independent sets by adjusting local detunings on the Rydberg Hamiltonian according to each vertex's vertex support, which is a quantity that represents connectivity between vertices. By doing so, the strategy successfully reduces the error rate three times for the checkerboard graphs with defects when the adiabaticity is sufficient. In addition, the strategy decreases the error rate for random graphs even when the adiabaticity is relatively insufficient. Moreover, it is shown that the strategy helps to prepare a quantum many‐body ground state by raising the fidelity between the evolved quantum state and a 2D cat state on a square lattice. Finally, the strategy is combined with the non‐abelian adiabatic mixing and this approach is highly successful in finding maximum independent sets compared to the conventional adiabatic evolution with local detunings.
\end{abstract}
\maketitle

\section{Introduction}
The quantum adiabatic algorithm (QAA) stands out as one of the most promising quantum algorithms on various platforms~\cite{RMM+09, QLR+09, SSW+21, OLK+19} due to its simplicity~\cite{AL18}. It leverages the adiabatic theorem to find a ground state of a target Hamiltonian by gradually evolving from the ground state of an easily solvable initial Hamiltonian~\cite{RC02,FGG02,EWL+21,WYW20}.

Interestingly, solutions to combinatorial optimization problems can be obtained by setting the target Hamiltonian as the cost function for these problems~\cite{HP00,CFGG02}. While these problems have practical real-world applications, many of them fall within the NP-hard class, for which no known classical polynomial-time algorithms exist~\cite{S05,FGG+01}. In this context, the QAA has garnered attention as a potential candidate for achieving quantum advantage~\cite{MBM+23}. 

Specifically, the QAA offers a significant advantage in terms of scalability where the cost function of the problems naturally resembles the hardware's Hamiltonian~\cite{EWL+21,SSW+21}. One of the most prominent examples of this alignment in combinatorial optimization problems is the unit-disk maximum independent set (MIS) problem on Rydberg atom arrays due to their intrinsic resemblance~\cite{PWZCL18,PWZCL+18}. This example serves as the primary focus of our work.

The MIS is defined as the largest subset of vertices in a graph such that no two vertices in the subset are adjacent. The problem of finding MIS is NP-hard in the worst case~\cite{B03}. The MIS has various practical applications~\cite{WLG+22}, including wireless networks~\cite{WAJ10} and large data sets~\cite{B03}. An important subclass of the MIS is MIS on a unit-disk graph, where finding it is also NP-hard in the worst case~\cite{CCJL90}. Given the distance-dependent nature of the Rydberg Blockade mechanism \cite{WLT+21}, only the unit-disk MIS problem can be directly encoded into a 2D Rydberg Hamiltonian \cite{PWZCL18}.


Multiple studies have been conducted to encode arbitrary connected graphs into unit-disk graphs to overcome this unit-disk limitation. In Ref. \cite{KKH+22}, the authors presented a protocol called the ``Rydberg quantum wire scheme" to connect two distant vertices by adding ancilla qubits in 3D space, which was experimentally tested. Also, Ref. \cite{NLW+23} suggested a method to construct 2D gadgets consisting of $O(N^2)$ ancilla qubits to enable arbitrary connectivity. Furthermore, Ref. \cite{GMOS23} proposed a scheme that utilizes local detunings to directly encode arbitrary connectivity at the Hamiltonian level and facilitate performance. Additionally, protocols to reduce the required time by introducing a new term to the Rydberg Hamiltonian have been proposed~\cite{SWM+23,CCL+23}. In Refs. \cite{YWW21,ZGY+23}, the authors revealed mathematical equivalence between non-abelian gauge matrix \cite{WZ84} and the Rydberg Hamiltonian, and proposed a new efficient adiabatic path motivated by this relationship. Notably, Ref. \cite{EKC+22} claimed to have experimentally demonstrated a superlinear speedup over the classical method for finding MIS on unit-disk graphs of up to 289 qubits using Rydberg atom arrays. However, this result was rebutted by Ref.~\cite{ASM+23}, which proposed a new classical strategy that outperformed the quantum method presented in Ref.~\cite{EKC+22}.

Nevertheless, an approximation scheme explicitly designed for the MIS problem on the Rydberg Hamiltonian does not currently exist. Approximating MIS for general graphs is a challenging task, and the best polynomial-time algorithms can only find approximate solutions within a polynomial fraction of the exact solution placing them in the APX-complete class~\cite{B03}. For unit-disk graphs, approximate solutions within a desired fraction of the exact solutions can be obtained in polynomial time (falling into the PTAS class). However, practical limitations arise when trying to achieve a valuable fraction of the exact solutions in this case~\cite{DDK+15}. Therefore, there is a need for a more efficient approximation scheme specifically tailored to Rydberg atom arrays.

In this study, we present a strategy to find approximate solutions to the MIS problem on a unit-disk graph by utilizing local detunings of the Rydberg Hamiltonian during the QAA. Our approach incorporates the ``vertex support" concept introduced in Ref.~\cite{BSK10} to estimate the likelihood of each vertex being included in MIS. These estimated potentials are assigned to each atom's detuning magnitude during the adiabatic evolution. This process encourages atoms with a higher likelihood of being included in MIS to end up in the Rydberg state at the conclusion of the evolution. We optimize the acceptance degree of vertex support in our strategy and compare its performance to the case with a conventional QAA scheme without local detunings. In the context of QAA, considering vertex support successfully reduces error rates for two types of graphs: checkerboard graphs with defects and random graphs. For checkboard graphs with defects, the error rate was reduced threefold when the adiabaticity was enough. For random graphs with a density of 3.0, our strategy reduced the error rate even when the adiabaticity was insufficient. This implies that our strategy is applicable to larger graphs, which are hard to deal with classical computers. Notably, our strategy can be applied to any type of encoding, as it does not rely on the specific characteristics of unit-disk graphs.

Following the above process, we can gain insights into the ground states of the given Hamiltonian. However, we remain uncertain whether the quantum state of our atoms is actually in the ground state. This uncertainty arises because a lack of information about the ground states necessitates classical post-processing after the evolution of our process. On the other hand, if the ground state of certain special system is known \cite{BSK+17,EWL+21,SSW+21}, we can maximize the probability of obtaining this ground state by harnessing the vertex support concept as a measure of connectivity. By doing so, we increase the fidelity of a 2D cat state \cite{OLK+19} from 0.68 to 0.90, which is a quantum superposition of two opposite states.

In addition, we have applied our strategy to the non-abelian adiabatic mixing \cite{YWW21,ZGY+23}. In the perspective of non-abelian adiabatic mixing, other waveforms $\Omega_i$ and $\phi_i$ should be different for each qubit. We compare this approach's performance with the conventional waveforms. By doing so, we obtain much better results when we use the gauge field approach. This implies that the adiabatic evolution for finding MIS with local detunings $\Delta_i$ is fully completed with consideration of the gauge field.

\section{Rydberg Hamiltonian and MIS cost function}
\label{sec:Rydberg}

\begin{figure}
	\centering
	\begin{subfigure}{0.49\linewidth}
		\centering
		\includegraphics[width=\textwidth]{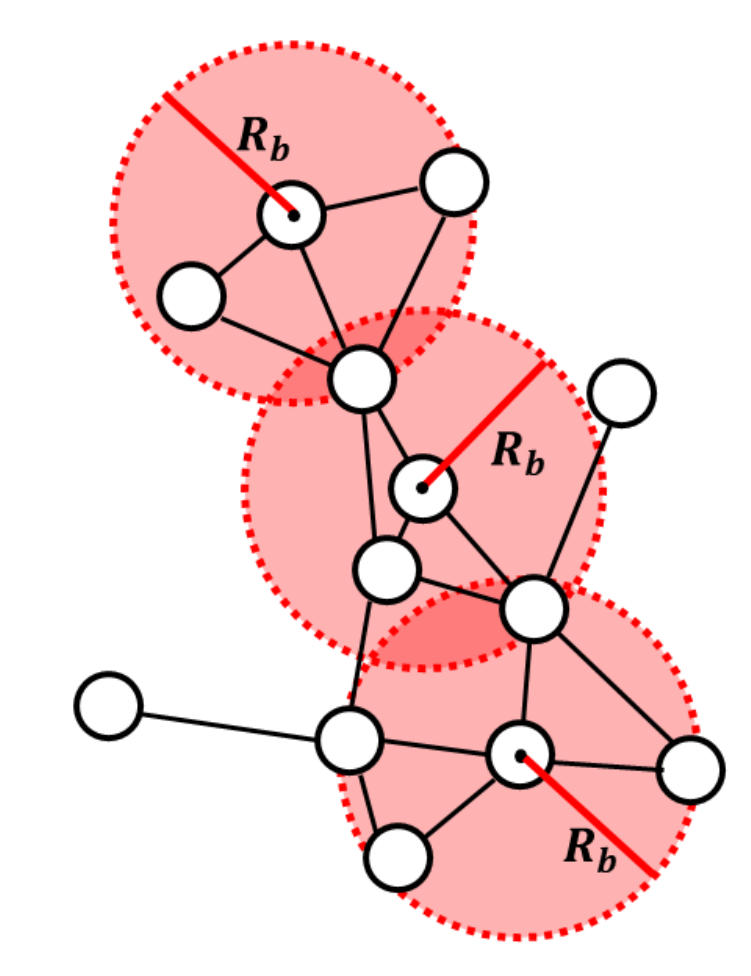} 
		\caption{}
		\label{fig:rydberg_blockade1}
	\end{subfigure}
	\hfill
	\begin{subfigure}{0.49\linewidth}
		\centering
		\includegraphics[width=\textwidth]{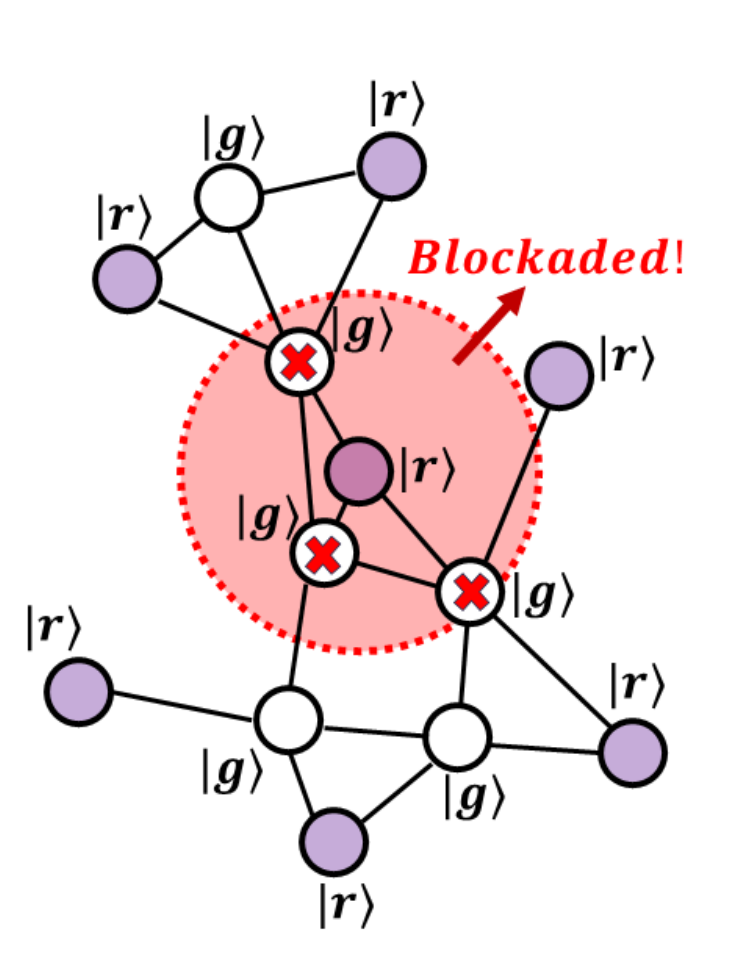}
		\caption{}
		\label{fig:rydberg_blockade2}
	\end{subfigure}
	\caption{a) Two vertices are connected when one vertex is within the Rydberg blockade radius $R_b$ of another, which allows us to regard the configuration of atoms as the unit-disk graph. The red circles denote the Rydberg blockade range of an atom. b) Only one atom can be excited to a Rydberg state in a Rydberg blockade radius by the Rydberg blockade mechanism. The color of Rydberg state atoms is purple, and the ground state atom is white.}
	\label{fig:rydberg_blockade}
\end{figure}

Consider a system consisting of $N$ atoms which satisfy the Rydberg Hamiltonian. For the Rydberg Hamiltonian, let us denote the $i$-th Rydberg atom array's ground state as $|g_i\rangle$, the Rydberg state as  $|r_i\rangle$, and the Rydberg number operator as $\hat{n}_i:=|r_i\rangle\langle r_i|$. The Rydberg Hamiltonian can be decomposed as follows,
\begin{equation}
	H_{\mathrm{Ryd}}=H_\Omega+H_\Delta+H_V
\end{equation}
where $H_{\Omega}$ is the driving part, $H_\Delta$ is the detuning part, and $H_{V}$ is the interaction part. The details of each Hamiltonian is as follows:
\begin{align}\label{drivinghamiltonian}
    & H_\Omega(t) = \sum_{j=1}^N \cfrac{\Omega_j(t)}{2}(e^{i\phi_j(t)}|g_j\rangle\langle r_j|+e^{-i\phi_j(t)}|r_j\rangle\langle g_j|), \\
    \label{detuninghamiltonian}
    & H_\Delta(t) = -\sum_{j=1}^N \Delta_j(t)\hat{n}_j, \quad \tn{and} \\
    \label{interactionhamiltonian}
    & H_V = \sum_{j<k}V_{jk}\hat{n}_j\hat{n}_k.
\end{align}
The symbols $\Omega_j$, $\phi_j$, and $\Delta_j$ characterize the Rabi frequency, the phase of the laser, and the driving laser field's detuning of the \textit{j}-th atom, respectively. Moreover, $V_{jk}=C_6/|\vec{r}_j-\vec{r}_k|^6$ is the Rydberg interaction, a type of van der Waals interaction between two Rydberg states where $C_6$ represents the interaction constant. We set $C_6$ as $2\pi \times 862690~\text{MHz}~\mu m^6$ in this work~\cite{WLT+21}. This intensive interaction creates the Rydberg blockade phenomenon, which restricts the excitation of more than one atom within a specific range. 

For the cost function of the MIS problem \cite{B03}, let us $G=(V, E)$ be a given undirected graph with $V$ and $E$ as the set of vertices and set of edges, respectively. Then, an independent set $I$ is defined as a subset of $V$ in which any two vertices in $I$ are not connected. The MIS is the independent set of which cardinality is maximum, i.e., $I_{\tn{MIS}}=\max_{|I|}(I)$. This statement is equivalent to minimizing the following cost function, which depends on a $S\subseteq V$, and it is given by
\begin{equation}
	\label{miscostfunction}
	H(S)=-\sum_{i\in V}\Delta_i n_i+\sum_{(i,j)\in E}U_{i,j}n_in_j,
\end{equation}
where
\begin{align}
	n_i=
	\begin{cases}
		1 & \quad \tn{if} \quad i\in S\\
		0 & \quad \tn{if} \quad i\notin S
	\end{cases}.
\end{align} 
Note that the symbol $\Delta_i$ indicates the assigned weights of each vertex. The MIS problem refers to the situation where all $\Delta_i$ are equal and the maximum weight independent set (MWIS) problem is when each $\Delta_i$ can differ. The MWIS is the independent set with a maximum summation of its vertices' weights. The symbol $U_{i,j}$ is the interaction strength between two vertices, where the condition $U_{i,j}>\max_k(\Delta_k)$ results in the independence constraint. The unit-disk graph is the graph where two vertices are connected by an edge if they are within a pre-defined unit-disk radius $d$, i.e., $|\overrightarrow{r_i}-\overrightarrow{r_j}|<d$ \cite{CCJL90}.  Figure~\ref{fig:rydberg_blockade1} describes the unit-disk graph with the unit-disk radius $R_b$.

The explicit similarity between Eq.~(\ref{miscostfunction}) and $H_\Delta+H_V$ reveals that the Rydberg atom arrays can be an efficient MIS problem solver by regarding a set of Rydberg states in the ground states as a maximum independent set. In fact, the cost function and its solutions to the MIS problem on unit-disk graphs can be efficiently encoded to the Rydberg Hamiltonian and its ground states due to the distance dependency of the Rydberg interaction ~\cite{PWZCL+18}, which is showcased in Figure \ref{fig:rydberg_blockade2}.  The only difference is that the interaction between two Rydberg states works at every distance, whether they are connected in the original graph. The Rydberg blockade radius $R_b$, corresponding to the unit-disk radius in a unit-disk graph, is conventionally defined as follow~\cite{LBR+16}.:
\begin{equation}\label{rydergblockaderadius}
    R_b:=\left(\cfrac{C_6}{\Omega}\right)^{1/6}
\end{equation}
Even though a potential term exists beyond $R_b$, it was demonstrated that almost every ground state of checkerboard graphs with defects is an MIS, and this long interaction term facilitates adiabatic process~\cite{EKC+22}. 

\section{Quantum adiabatic algorithm}
\label{sec:QAA}
\begin{figure}
\centering
\begin{subfigure}{0.49\linewidth}
  \centering
  \includegraphics[width=\textwidth]{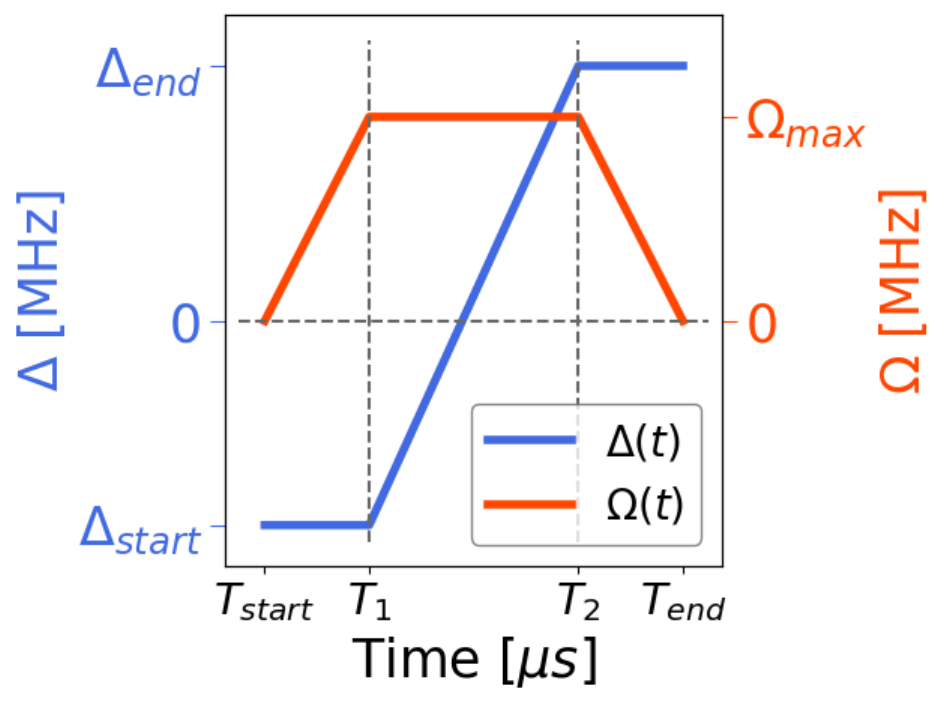} 
  \caption{}
  \label{fig:waveform}
\end{subfigure}
\hfill
\begin{subfigure}{0.49\linewidth}
  \centering
  \includegraphics[width=\textwidth]{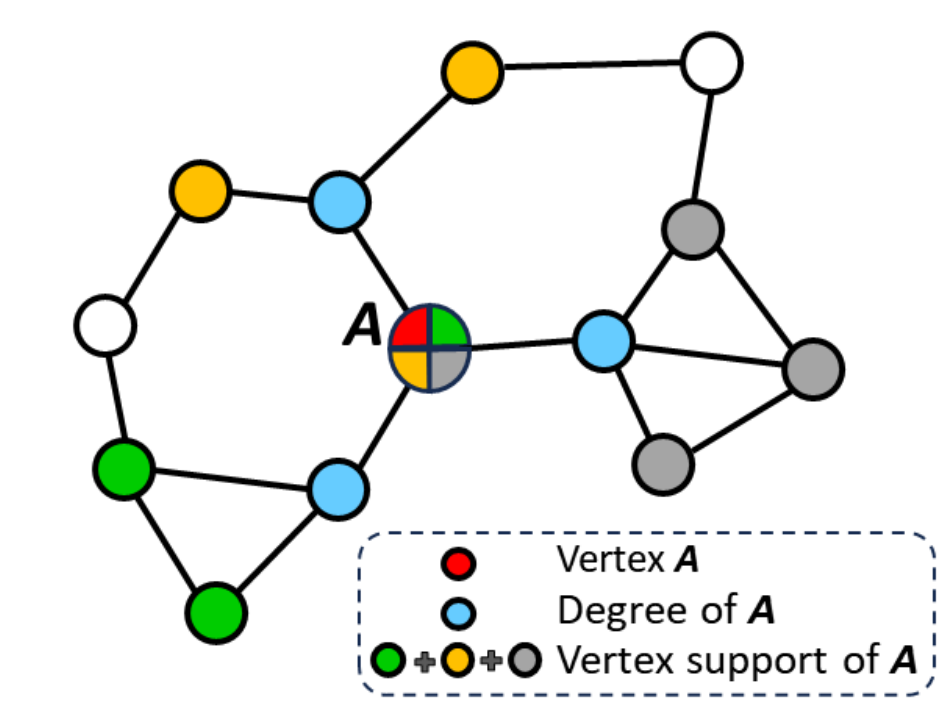}
  \caption{}
  \label{fig:vertex_support}
\end{subfigure}
\caption{a) Waveforms of $\Omega(t)$ and $\Delta(t)$ employed on adiabatic evolution. b) Vertex support of \textit{A}.}
\label{fig2}
\end{figure}

There are mainly two strategies to obtain ground states of the target Hamiltonian: adiabatic approaches and diabatic approaches, which include quantum approximated optimization algorithm (QAOA) \cite{FGG14,ZWC+20}. The primary focus of this paper is on the former approach, the so-called QAA. Notably, it was reported that QAA performs better than QAOA in finding MIS using the Rydberg atom arrays~\cite{EKC+22}. This section discusses how to apply the QAA to the Rydberg atom arrays for finding MIS. The overall waveforms are in Figure~\ref{fig:waveform}. To focus on the effect of our strategy, We set $\Omega_i=\Omega$ and $\phi_i=0$ for all $i$.

The QAA uses the adiabatic theorem to prepare the desired quantum many-body ground states. The adiabatic theorem guarantees that a system remains in its initial eigenstate if its evolution is sufficiently slow. In this paper, the QAA begins by preparing the state $\bigotimes_{i=1}^N |g_i\rangle$ which is the ground state of $H_{\mathrm{Ryd}}(t=T_\mathrm{start})$ where $\Omega=0$ and all $\Delta_i<0$. The reason why we prepare $\bigotimes_{i=0}^N |g_i\rangle$, not the conventional initial state $|+\rangle^{\otimes N}= \bigotimes_{i=1}^N (|g_i\rangle+|r_i\rangle)/\sqrt{2}$) is that the presence of the interaction Hamiltonian $H_V$ makes it hard to prepare $|+\rangle^{\otimes N}$. The next step is to evolve the initial state $\bigotimes_{i=1}^N |g_i\rangle$ to find a path to the solution space. This evolution is achieved by gradually increasing $\Omega$ to its maximum value $\Omega_{\max}$. Simultaneously, $\Delta_i$ is increased until it reaches $\Delta_{i,\mathrm{end}}$. Finally, when all $\Delta_i$ values have reached their maximum value, $\Omega$ is reduced to $0$, making the system's Hamiltonian $H_{\mathrm{Ryd}}(t=T_\mathrm{end})$ correspond to the MIS problem's cost function as described in Eq.~(\ref{miscostfunction}). 

We use a trapezoidal shape $\Omega$ and a piecewise linear shape $\Delta_i$ for simplicity. However, since one of the most important criteria for adiabatic evolution is the rate of change of the Hamiltonian over time \cite{A09,T10,KTF+19}, these waveforms can be smoothed by summing sinusoidal functions~\cite{OLK+19} or by breaking them into smaller segments and optimizing them. Specifically, according to Ref. \cite{AK21}, adiabaticity can be recovered to second order in the inverse of the total evolution time $O(1/T^2)$ if the rate of change of the Hamiltonian is continuous and zero at both the beginning and the end of the evolution. To demonstrate the effect of smoothing the Hamiltonian, we additionally exploit three waveforms, $\sin(\pi t),~\sin^2(\pi t),$ and $\sin^3(\pi t)$, to evolve quantum states without the classical post-processing. The results of applying these other waveforms are described in Appendix B. Additionally, Refs. \cite{YWW21,ZGY+23} suggested other smooth waveforms based on the relationship between the gauge field and the Rydberg Hamiltonian. From the perspective of adiabatic evolution using the gauge field, local detunings $\Delta_i$ imply that the rotation rates of individual qubits are different, so applying local values for $\Omega_i$ and $\phi_i$ is also necessary. The performance of these waveforms is discussed in Sec. \ref{non-abelian}.

Another key variable for maintaining adiabaticity is the minimum energy gap, $\delta_{\min}$, between a ground state and a first excited state that is coupled with the ground state during quantum evolution~\cite{RSS+13}. The time $T$ required to sustain adiabaticity should satisfy $T> 1/\delta^2_{\min}$~\cite{FGGS00}. However, for graphs with complex configurations, the long-range Rydberg potential and the large degeneracy in both the ground states and the first excited states make it difficult to determine the \emph{true} minimum energy gap. Additionally, the classical post-processing in our algorithm diminishes the role of $\delta_{\min}$ in maintaining adiabaticity. For these reasons, we can not use $\delta_{\min}$ as the primary metric for our strategy. Instead, we demonstrate that our strategy decreases $\delta_{\min}$ while evolving atoms that form a square lattice graph, which does not have ground state degeneracy. 

\section{Vertex support}
\label{sec:Vsupp}
The vertex support is a measure of how many vertices are around a particular vertex. It was proposed by Balaji \emph{et al}.~\cite{BSK10} as part of a heuristic approach for solving the MIS problem, named the vertex support algorithm. To define the vertex support, consider an undirected graph $G=(V,~E)$, where $V$ is the set of total vertex and $E$ is the set of total edges. The neighborhood of the \textit{i}-th vertex, denoted by $\mathcal{N}_i$, is the subset of $V$ and consists of all vertices that are connected to \textit{i}-th vertex. The degree of vertex \textit{i}, $d_i$, is the number of vertices in $\mathcal{N}_i$, i.e., $d_i=|\mathcal{N}_i|$. Then, the support of the \textit{i}-th vertex, $s_i$, is defined as the sum of degrees of all vertices in $\mathcal{N}_i$, i.e., $s_i=\sum_{j\in \mathcal{N}_i} d_j$. These definitions are described in Figure \ref{fig:vertex_support}. It is evident that a vertex with large vertex support interacts more extensively with other vertices than one with lower vertex support. For approximating MIS, we incorporate this point into the local detuning $\Delta_i$ in Eq.~(\ref{detuninghamiltonian}). 

\section{Adjusting local detunings according to vertex supports}
\label{sec:localdetun}
The primary assumption in this paper is that a vertex with more connections to others is less likely to be included in MIS. Note that this assumption is not restricted to the unit-disk structure, so other encoding schemes can also benefit from our strategy. To apply this assumption to the Rydberg Hamiltonian, we use vertex support as a measure of connectivity between vertices. While an atom with more connections to others is inherently less likely to be in a Rydberg state due to the strong Rydberg interaction in  Eq.~(\ref{interactionhamiltonian}), we explicitly account for connectivity by adjusting the local detuning $\Delta_i$ in Eq.~(\ref{detuninghamiltonian}) for each atom individually. This adjustment is accomplished by subtracting a scaled vertex support value from the $\Delta_{\mathrm{start}(\mathrm{end})}$. As a result, the Rydberg state's energy level alters according to the magnitude of each atom's vertex support. Also, the values of a vertex's degree and vertex support are quite similar in the graphs considered in this paper, so the choice between them does not significantly affect the final results. However, we anticipate that selecting vertex support could have a much greater impact on more complex and larger graphs, given the superior performance of the vertex support algorithm in the classical regime \cite{BSK10}.

In fact, assigning different values to each local detuning transforms the system's Hamiltonian from representing the MIS problem's cost function to representing the MWIS problem's cost function. Effectively, our approach can be conceptualized as a search for a weighted unit-disk graph whose MWIS is similar to the original graph's MIS. Due to the adjusted local detunings by our strategy, this MWIS is easier to obtain by the adiabatic process than the original maximum independent sets. 

To maximize the total Rydberg density $\langle n\rangle=\sum_{i\in V} \langle n_i\rangle$, we optimize a scaling factor of the vertex support $\epsilon$, $\Delta_{\mathrm{start}}$, and $\Delta_{\mathrm{end}}$ shown on Figure \ref{fig:waveform}. However, trials to maximize $\langle n\rangle$ can result in excessively enormous values of $\Delta_{\mathrm{end}}$, violating the independence constraint. Therefore, we employ two classical post-processing strategies: the vertex reduction and the vertex addition in Ref.~\cite{EKC+22}. After sampling the evolved quantum states, the vertex reduction removes pairs of vertices, violating the independence constraint. Then, the vertex addition appends vertices to the independent set, which does not cause independence violations. The summary of our strategy is as follows.

\begin{enumerate}
	\item For a given undirected graph $G=(V,~E)$, calculate a vertex support of each vertex $\{s_i\}_{i\in V}$ and a mean of the vertex supports,
	\begin{equation}
		\mathbb{E(\mathbf{s})}=\frac{1}{N}\sum_{i\in V} s_i.
	\end{equation}

	\item Set each local detuning $\Delta_i$ of the $i$-th atom as follows,
	\begin{equation}
		\label{weighted_hamiltonian}
		\Delta_{i,~\mathrm{start(end)}}:=\Delta_{\mathrm{start(end)}}+(\mathbb{E(\mathbf{s})}-s_i)\times \epsilon.
	\end{equation}
	We fixed several values throughout this paper: $\Omega_{\max}=2\pi\times 4$, $T_1=T_{\mathrm{end}}/7$, and $T_2=T_{\mathrm{end}}\times 6/7$ where $T_1$ and $T_2$ come from Figure \ref{fig:waveform}. We empirically find that the initial values $\Delta_{\mathrm{start}}=-2\pi\times 8~\mathrm{MHz},~\Delta_{\mathrm{end}}=2\pi\times 10~\mathrm{MHz}$ with $\epsilon=1.5$ generally provide good results for the classes of graphs in our paper.	
	\item After enough evolution, sample the quantum state by basis \{$|g_i\rangle,~|r_i\rangle$\}.	
	\item Exploit vertex reduction and vertex addition on the sampled results sequentially. Then, calculate the total Rydberg density $\langle n \rangle $.	
	\item Optimize $\Delta_{\mathrm{start}},~\Delta_{\mathrm{end}}$, and $\epsilon$ to maximize $\langle n\rangle$ until the desired optimization threshold is reached. More parameters can be added to deform the waveform of $\Delta(t)$~\cite{EKC+22,OLK+19}. 
\end{enumerate}

Note that because this strategy is based on the assumption mentioned at the beginning of this section, the sign of the optimized $\epsilon$ may be negative for some graphs, which contradicts the assumption. Nevertheless, these cases accounted for only about 2\% of all simulated graphs. Even in these cases, the performance in approximating the maximum independent set arises.

We obtain the value of the exact cost function from the evolved quantum states without noise to identify the true power of this strategy. The gradient-free Nelder-Mead optimizer~\cite{GH12} showed the best performance among several optimizers. Most of our numerical calculations were done with \texttt{Bloqade.jl}~\cite{Blo23}, an open-source package developed for simulating a Rydberg atom-based architecture.

\begin{figure}[h]
	\centering
	\begin{subfigure}{0.49\linewidth}
		\centering
		\includegraphics[width=\textwidth]{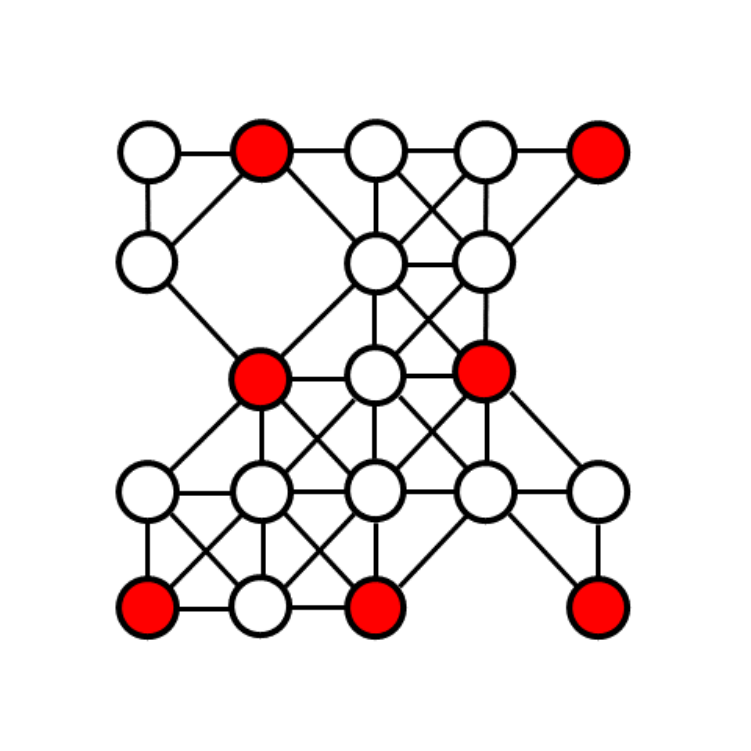} 
		\caption{}
		\label{fig:square_with_defects}
	\end{subfigure}
	\hfill
	\begin{subfigure}{0.49\linewidth}
		\centering
		\includegraphics[width=\textwidth]{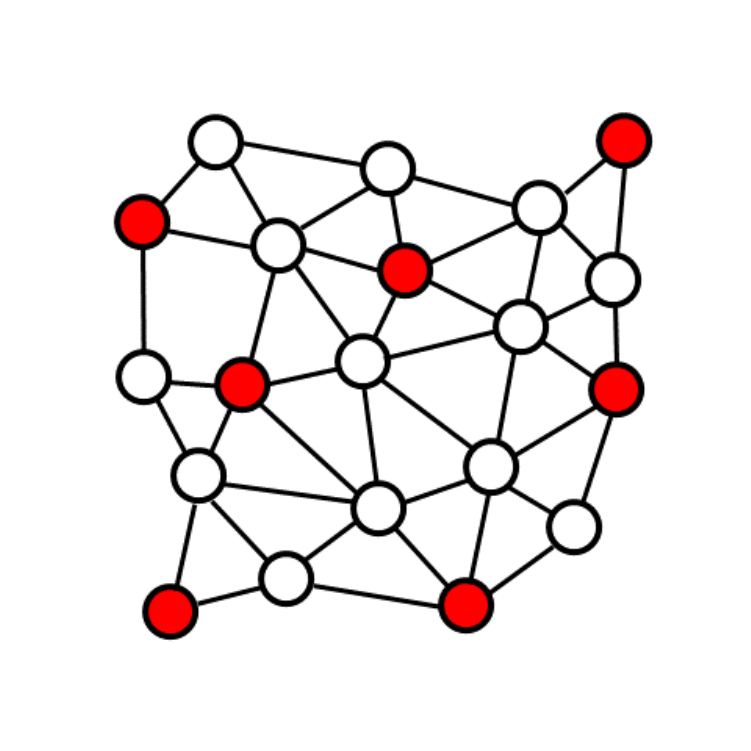}
		\caption{}
		\label{fig:random_graph_config}
	\end{subfigure}
	\caption{a) A checkerboard graph with defects probability 0.2 and its maximum independent set is colored red. b) A random graph of density of 3.0 and its maximum independent set is colored red.}
	\label{fig3}
\end{figure}

\section{Numerical simulations}
\label{sec:numsim}
In this section, we apply our strategy to two types of graphs, which are described in Figure~\ref{fig:square_with_defects} and Figure~\ref{fig:random_graph_config}. The first one is a checkerboard graph with defects, and the other one is a random graph generated by the procedure in Ref.~\cite{SMA20}.

\begin{figure*}[ht]
	\centering
	\begin{subfigure}{0.328\linewidth}
		\centering
		\includegraphics[width=\textwidth]{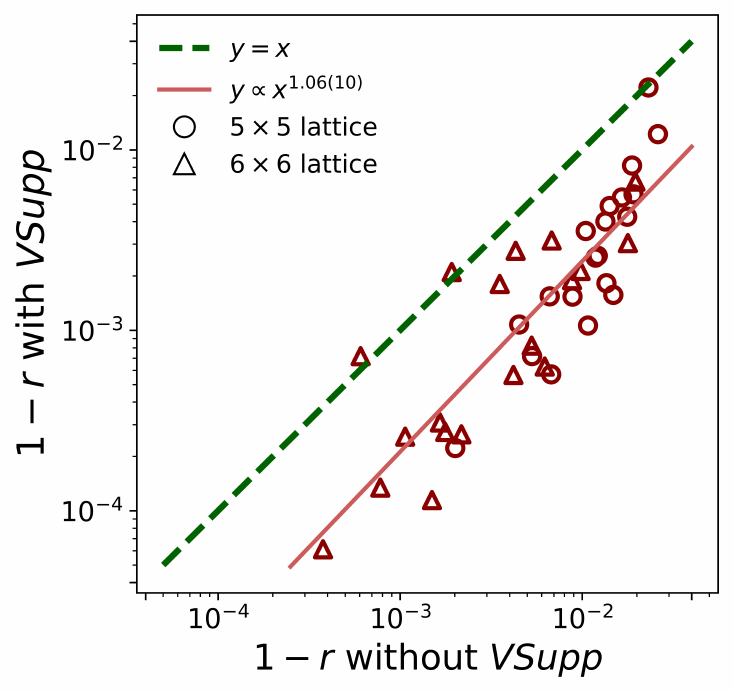} 
		\caption{}
		\label{fig:unionjack_1}
	\end{subfigure}
	\hfil
	\begin{subfigure}{0.328\linewidth}
		\centering
		\includegraphics[width=\textwidth]{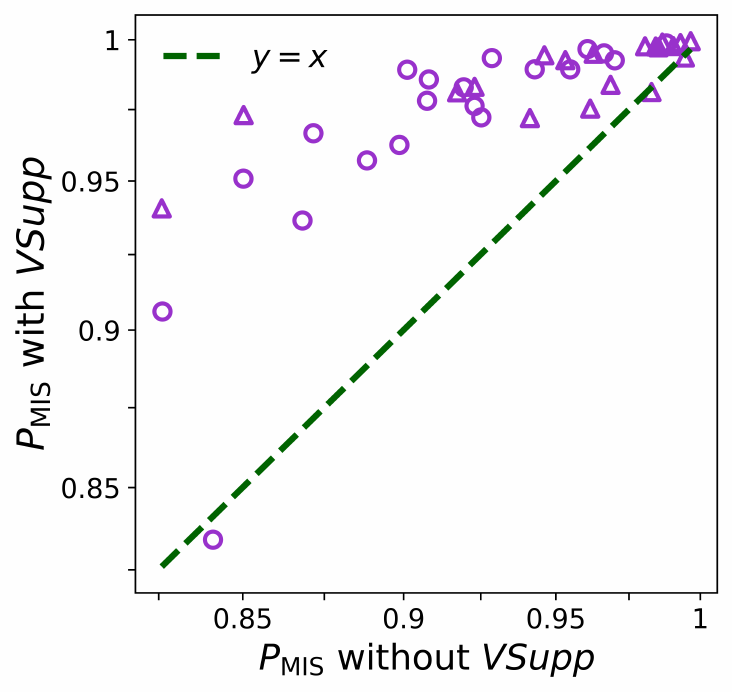}
		\caption{}
		\label{fig:unionjack_2}
	\end{subfigure}
	\hfil
	\begin{subfigure}{0.334\linewidth}
		\centering
		\includegraphics[width=\textwidth]{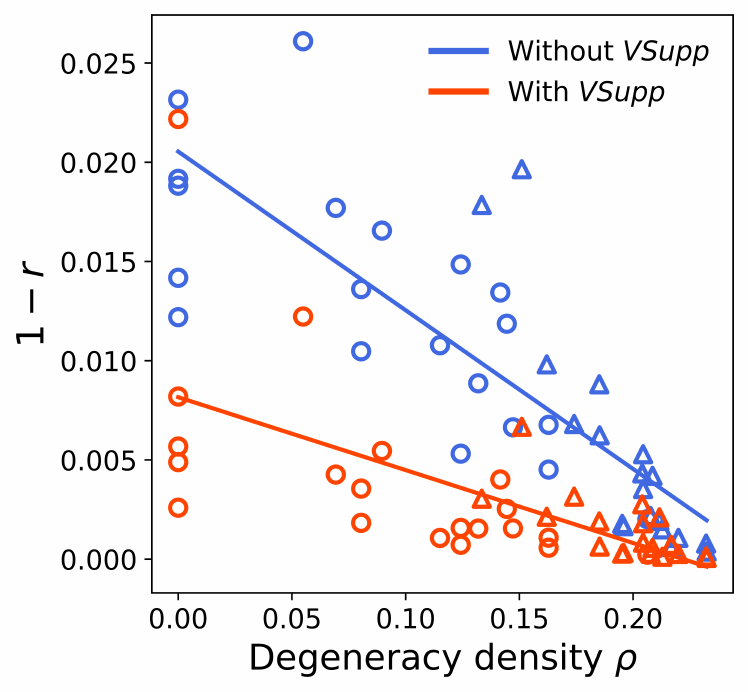}
		\caption{}
		\label{fig:unionjack_3}
	\end{subfigure}
	\caption{Optimized values after evolving randomly generated $5\times 5$ and $6\times 6$ checkerboard graphs with defects during $1~\mu s$, 20 instances for each size. The symbols to notate the size of graphs are showcased in Figure \ref{fig:unionjack_1}. a) The error rate $1-r$ `with \textit{VSupp}' versus `without \textit{VSupp}' is described by $\log$-$\log$ scale. The power law curve is employed to fit the results.  b) The probability of getting MIS with the strategy `without \textit{VSupp}' versus the probability of getting MIS with the strategy `with \textit{VSupp}' is described. c) Comparing error rates $1-r$ between `without \textit{VSupp}' and `with \textit{VSupp}' according to the degeneracy density $\rho$. Linear fit of `without \textit{VSupp}' and `with \textit{VSupp}' is also described together.} 
	\label{fig:unionjack}
\end{figure*}
To simulate a quantum evolution of atoms up to 29, which is intractable for classical computers, we neglect the Rydberg blockade violation of two atoms within a near range where the condition $V(|\vec{r}_j-\vec{r}_k|)\gg\Omega, \Delta$ is satisfied.

In the approximation scheme, we measure the performance of our strategy using two key metrics: approximation ratio $r$ and error rate $1-r$,
\begin{equation}
	r:=\frac{\langle n\rangle}{|I_{\tn{MIS}}|},
\end{equation}
where the symbol $\langle n\rangle$ and $|I_{\tn{MIS}}|$ denotes the total Rydberg density and the cardinality of the MIS,  respectively. It was suggested that the error rate $1-r$ and the degeneracy density $\rho=\log{(D_{|I_{\tn{MIS}}|})}/N$ is linearly dependent for the checkerboard graph with defects, where the $D_{|I_{\tn{MIS}}|}$ is the degeneracy of the maximum independent sets~\cite{EKC+22}. In the rest of this paper, we will refer our strategy as `with \textit{VSupp}' which optimizes $\{\Delta_{\mathrm{start}}, \Delta_{\mathrm{end}}, \epsilon\}$ and the conventional QAA without local detunings as `without \textit{VSupp}' which optimizes $\{\Delta_{\mathrm{start}},\Delta_{\mathrm{end}}\}$ where both strategies encompass the classical post-processing.

The checkerboard graphs in this study connect nearest neighbor pairs and next nearest neighbor pairs. This next nearest connection makes us deal with non-planar graphs by Rydberg atom arrays. We set the nearest distance between two atoms as $4.5~\mu m$ where the Rydberg blockade radius $R_b\approx 7.74~\mu m$ by Eq.~(\ref{rydergblockaderadius}). We neglect the Rydberg blockade violation between nearest neighbors given that $C_6/(4.5~\mu m)^6\approx 25.9\Omega \gg \Omega, \Delta$. 20\% of total atoms in graphs are removed randomly to create defects.  

After $1~\mu s$ time evolution of checkerboard graphs, in Figure \ref{fig:unionjack_1}, the power-law fitting between the error rate of `without \textit{VSupp}' and `with \textit{VSupp}' is $x^{1.06(10)}/3.16$, which represents high linearity between them. This evolution has sufficient adiabaticity because the probability of obtaining maximum independent sets, $P_{\tn{MIS}}$, is around 0.9 in Figure~\ref{fig:unionjack_2}. According to these figures, our strategy linearly reduces the error rate by three times for checkerboard graphs with defects when adiabaticity is sufficient. Notably, in Figure~\ref{fig:unionjack_3}, the error rate of both `with \textit{VSupp}' and `without \textit{VSupp}' shows weak linear dependency on the degeneracy density $\rho$, unlike the error rate of Ref.~\cite{EKC+22}. This relatively weak linear dependency implies that the finite size effect plays a role due to the small size of our $5\times 5$ and $~6\times 6$ graphs, unlike the size of graphs in Ref.~\cite{EKC+22}, which is up to 289 vertices. Furthermore, the cardinality of MIS in these graphs is 7 to 10, which is too small that the cardinality, rather than the degeneracy density $\rho$, might affect the error rate. Even though the tendency between the error rate and the degeneracy density is well conserved, verifying our strategy on larger graphs in actual experiments is necessary to clarify our strategy's performance. 

In addition, to ensure that our strategy helps to prepare low-energy states, we directly optimize $P_{\textrm{MIS}}$ without the classical post-processing, not $\langle n \rangle$. The result demonstrates that the instances with high $P_{\textrm{MIS}}$ without the classical post-processing have a tendency to also get high $P_{\textrm{MIS}}$ with the classical post-processing. Relevent plots are described in Appendix B.

The random graphs, which are more general in practical problems, are characterized by the number of atoms and their density. Following the generation process in Ref.~\cite{SMA20}, we generated random graphs with an exclusion radius of  $4~\mu m$ for experimental realization and a density of 3.0 to make it possible to simulate a large number of atoms up to 28 with appropriate approximation. Also, we dismiss the Rydberg blockade violation inside $4\sqrt{2}~\mu m$ where $C_6/(4\sqrt{2}~\mu m)^6 \approx 6.6\Omega > \Omega,\Delta$. 

\begin{figure*}
	\centering	
	\begin{subfigure}{0.33\linewidth}
		\centering
		\includegraphics[width=\textwidth]{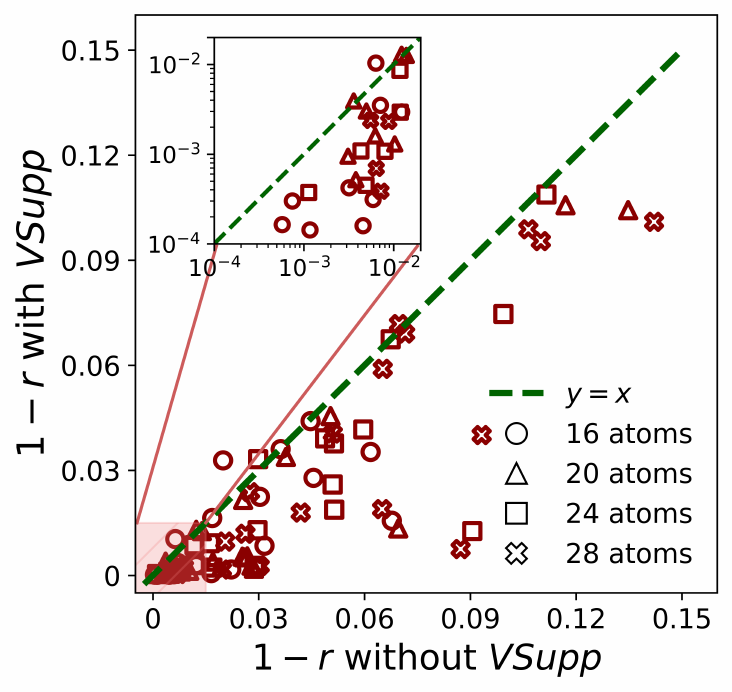}
		\caption{}
		\label{fig:random_graph_1}
	\end{subfigure}
		\hfil
	\begin{subfigure}{0.33\linewidth}
		\centering
		\includegraphics[width=\textwidth]{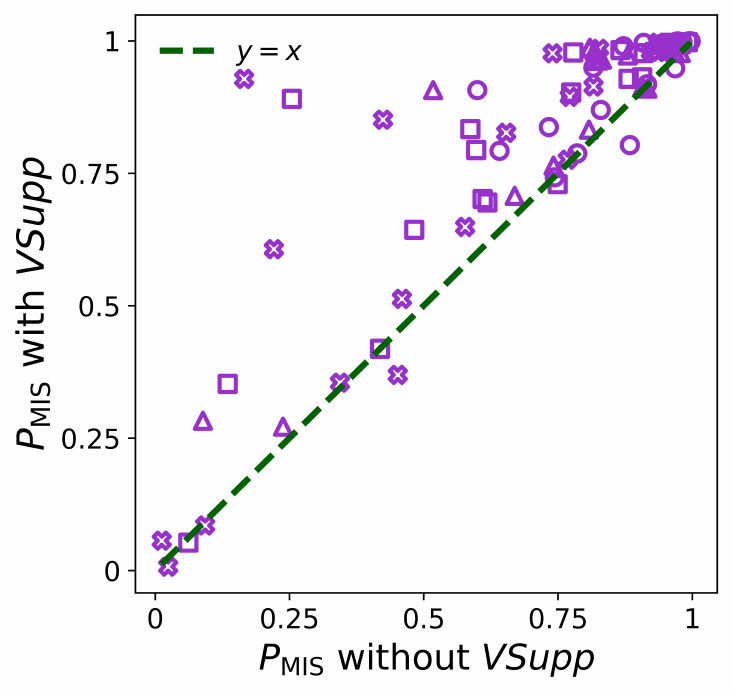} 
		\caption{}
		\label{fig:random_graph_2}
	\end{subfigure}
		\hfil
	\begin{subfigure}{0.33\linewidth}
		\centering
		\includegraphics[width=\textwidth]{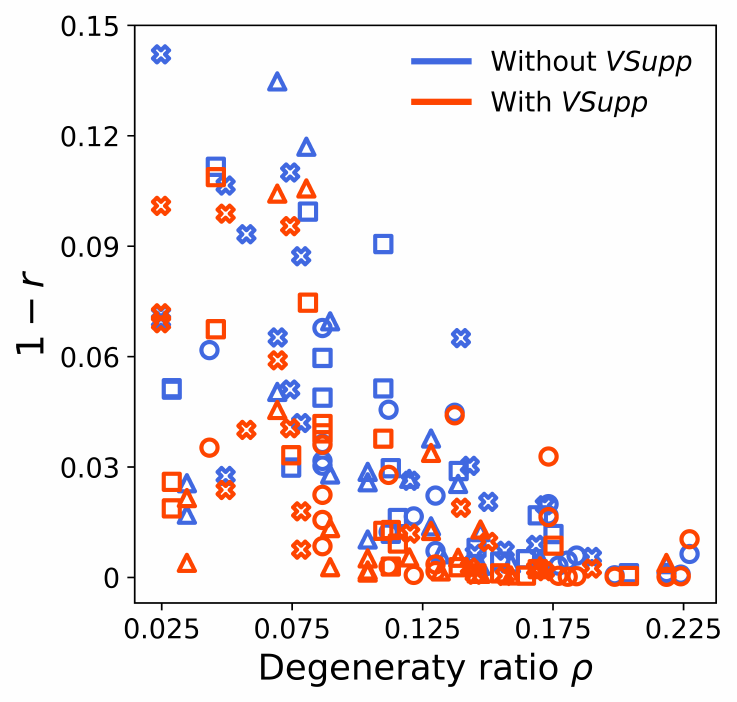}
		\caption{}
		\label{fig:random_graph_3}
	\end{subfigure}
	\caption{Optimized values after evolving random graphs of the density of 3.0 during $1~\mu s$. Twenty instances are simulated for each graph with 16, 20, 24, and 28 atoms.  The symbols to notate the size of graphs are showcased in Figure \ref{fig:random_graph_1}. a) The error rate $1-r$ `without \textit{VSupp}' versus `with \textit{VSupp}' is described. The inset figure magnifies the area between $10^{-4} \leq x,y \leq 2\times 10^{-2}$ and utilizes $\log$-$\log$ scale. b) The probability of getting MIS with the strategy `without \textit{VSupp}' versus the case of `with \textit{VSupp}' is described. c) Comparing error rates $1-r$ between `without \textit{VSupp}' and the case of 'with \textit{VSupp}' according to the degeneracy density $\rho$.}
	\label{fig:random_graph}
\end{figure*}

\begin{figure*}
	\centering	
	\begin{subfigure}{0.324\linewidth}
		\centering
		\includegraphics[width=\textwidth]{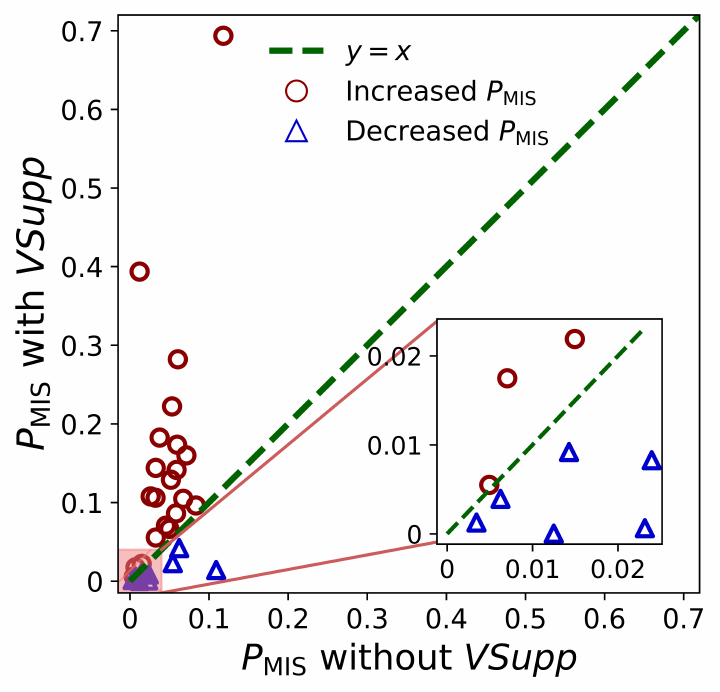}
		\caption{}
		\label{fig:randomhard_1}
	\end{subfigure}	
	\begin{subfigure}{0.324\linewidth}
		\centering
		\includegraphics[width=\textwidth]{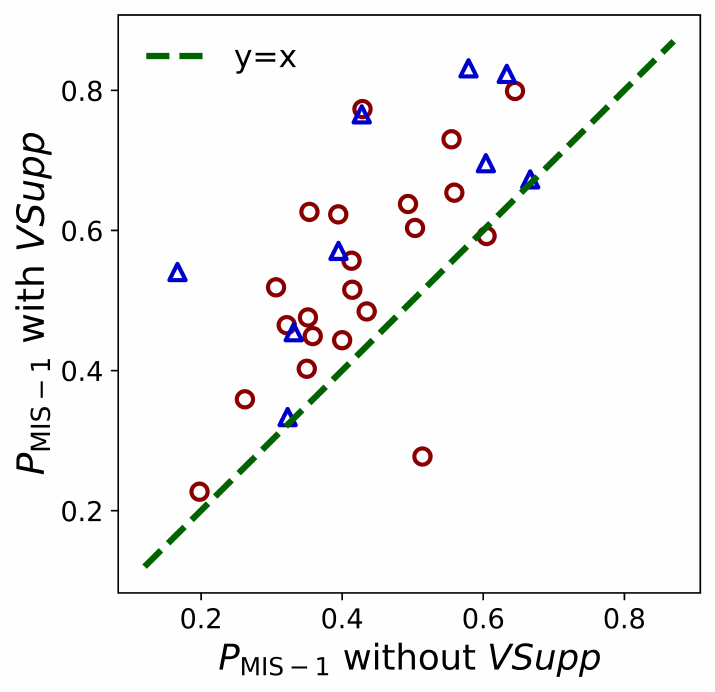} 
		\caption{}
		\label{fig:randomhard_2}
	\end{subfigure}
	\hfil
	\begin{subfigure}{0.342\linewidth}
		\centering
		\includegraphics[width=\textwidth]{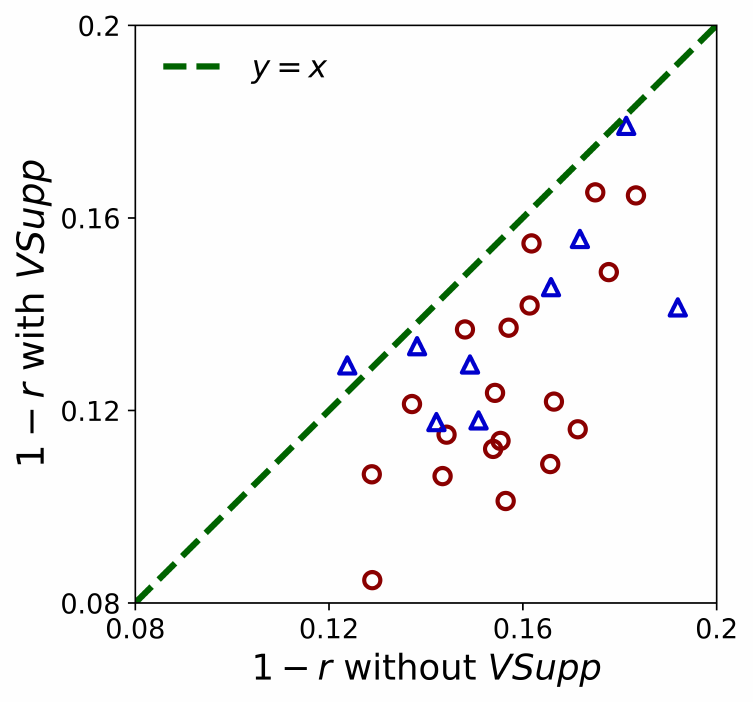}
		\caption{}
		\label{fig:randomhard_3}
	\end{subfigure}	
	\caption{Optimized values after evolving random graphs of the density of 3.0, ground degeneracy $D_{\tn{MIS}}=1$, during $0.4~\mu s$. Thirty instances are simulated where the number of vertices is 28. The red circle indicates the instances in which $P_{\tn{MIS}}$ are increased, and the blue triangle indicates the instances in which $P_{\tn{MIS}}$ are decreased. a) The probability of getting MIS with the strategy `without \textit{VSupp}' versus $P_{\tn{MIS}}$ with the strategy `with \textit{VSupp}' is described. The inset figure magnifies the area between $0 \leq x,y \leq 0.2$. b) The probability of getting the first excited state, which is named as $P_{\tn{MIS}-1}$, with the strategy `without \textit{VSupp}' versus the case of 'with \textit{VSupp}' is described. c) The error rate $1-r$ with the strategy `without \textit{VSupp}' versus the case of `with \textit{VSupp}' is described.}
	\label{fig:randomhard}
\end{figure*}

After $1~\mu s$ time evolution of several random graphs, points in Figure~\ref{fig:random_graph_1} are dominantly distributed under the line $y=x$ through $10^{-4}\leq x \leq 10^{-1}$, which exhibits the success of our strategy. Some instances show a dramatic decrease in the error rate even for the high error rate with the strategy `without \textit{VSupp}.' Although the evolution time is identical and the range of degeneracy ratio is similar to that of Figure~\ref{fig:unionjack}, the range of error rate and the $P_{\tn{MIS}}$ is relatively more expansive than that of Figure~\ref{fig:unionjack} which shows that the random graphs of the density of 3.0 are generally complex to solve than the checkerboard graphs with defects. In conclusion, 92.5\% of the total instances' error rate is decreased by considering the vertex support. Both `without \textit{VSupp}' and `with \textit{VSupp}' indicate weaker linear dependency on the degeneracy ratio than the dependency in Figure~\ref{fig:unionjack_3}.

To demonstrate the performance of our strategy in classically intractable larger graphs, we apply our strategy to more challenging graphs with shorter time, which results in less adiabaticity during evolution. In Figure~\ref{fig:randomhard}, 30 random graphs with ground degeneracy $D_{|I_\tn{MIS}|}=0$ are chosen because there is a tendency that the graphs with small degeneracy ratio $\rho$ have relatively large error rate $1-r$ on Figure~\ref{fig:unionjack_3}, \ref{fig:random_graph_3}.

With less adiabaticity, the $P_{\tn{MIS}}$ decreased for 9 out of 30 random graphs, which is described in Figure~\ref{fig:randomhard_1}, where $P_{\tn{MIS}}$ of `without \textit{VSupp}' is mostly under $10^{-1}$. However, in Figure~\ref{fig:randomhard_2}, our strategy increased the probability of getting the first excited state, $P_{\tn{MIS}-1}$, for almost every graph. As a result, the mean error rate of `without \textit{VSupp}', $0.16\pm0.02$, reduces to $0.12\pm0.03$ by our strategy in Figure~\ref{fig:randomhard_3}. Notably, the error rate is decreased even for the graphs in which $P_{\tn{MIS}}$ is decreased. This decrease implies that our strategy can be an excellent suggestion to approximate more challenging graphs, which we cannot sustain sufficient adiabaticity with current technology.

\section{Preparing a 2D cat state}
\label{sec:2dcat}
Although we find MIS through adiabatic evolution, there is uncertainty about obtaining a \emph{physical quantum ground state} after the above scheme because of the classical post-processing after measurements. This classical post-processing is essential to obtain MIS in our process, as maximizing the total Rydberg density $\langle n\rangle$ by optimizing $\Delta_{\mathrm{end}}$ without the classical post-processing may lead all atoms to become the Rydberg state. 

\begin{figure}[h]
	\centering
	\includegraphics[width=0.69\linewidth]{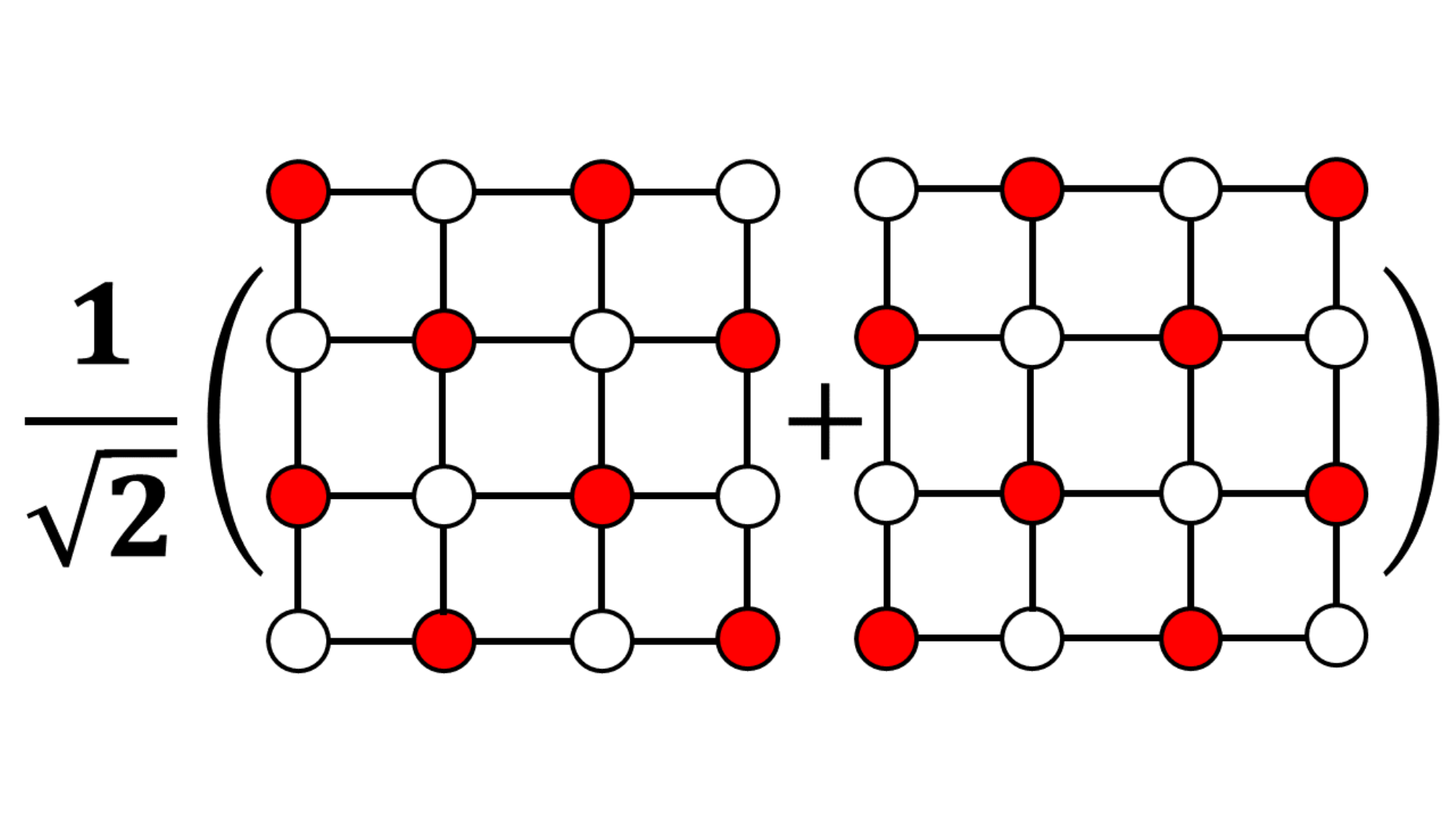}	
	\caption{Cat state on a $4\times 4$ square lattice where nearest neighbor vertices are connected.}
	\label{fig:catstate}
\end{figure}

Specifically, a superimposed ground state is hard to prepare directly by adjusting local detunings, not like computational basis states. Our scheme is applicable for this case; the desired ground state is a superimposed state and advantageous in various contexts. For example, the cat state is a superposition of two opposed states, such as $|000\dotsi 0\rangle$ and $|111\dotsi 1\rangle$, which can be used to benchmark quantum hardware~\cite{OLK+19,SXL+17,SUP+17}, quantum error correction~\cite{SDM+22}, and quantum network~\cite{HBW+19}. For the Rydberg atom arrays, the 1D cat state $(|010\dotsi 1\rangle + |101\dotsi 0\rangle)/\sqrt{2}$ is experimentally prepared up to 20 atoms with fidelity $F=\langle \psi_{\tn{CAT}}|\rho|\psi_{\tn{CAT}}\rangle \geq 0.5$ by applying local detunings at the edges of configuration~\cite{OLK+19}. In this section, we prepare the 2D cat state described in Figure \ref{fig:catstate} by applying local detunings according to the vertex support as a concept to quantify connectivity. A $4\times 4$ square lattice was used to demonstrate the increase in fidelity achieved by our strategy. Notably, there was no approximation during numerical quantum evolution, as the distance between two adjacent atoms was sufficiently large.

\begin{figure}[h!]
	\centering
	\includegraphics[width=0.69\linewidth]{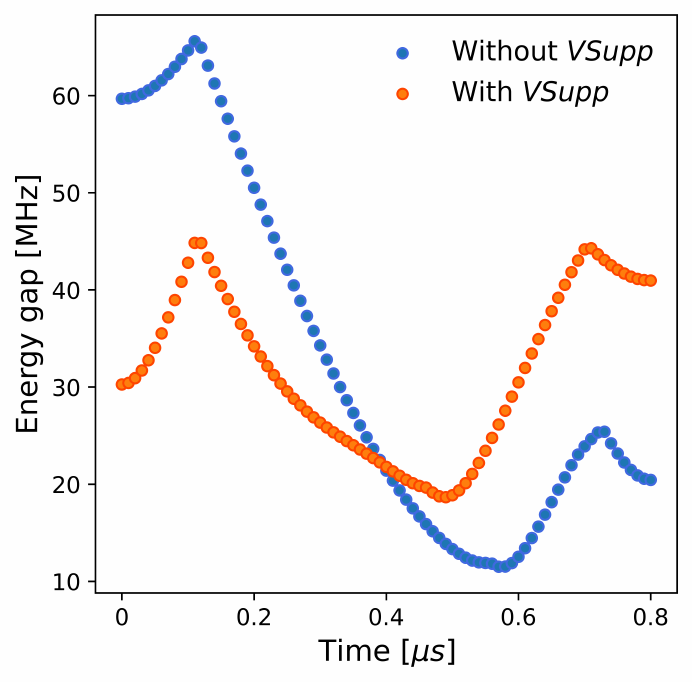}	
	\caption{Comparing the energy gap between the ground state and the coupled first excited state of `without \textit{VSupp}' and `with \textit{VSupp}' according to evolution time.}
	\label{fig:energy_gap}
\end{figure}

During $0.8~\mu s$ time evolution, the energy gap between the ground state and the coupled first excited state is plotted in Figure~\ref{fig:energy_gap}. Our strategy increases the minimum energy gap $\delta_{\min}$ from 11.5 to 18.7. This enhanced adiabaticity influences the fidelity between the cat state and the evolved quantum state. The fidelity between them increases from 0.68 to 0.90. Notably, the optimized $\epsilon$ multiplied by the vertex support is negative, which is opposed to the default sign. This negative sign is because, for the square lattice in Figure~\ref{fig:catstate}, the vertices at the edges are more likely to be in a Rydberg state than those with more connection with others. This consequence suggests that whether or not our assumption can not precisely recommend the ground states, its ability to quantify connectivity works successfully. The vertex support quantifies connectivity well and aids in preparing specific quantum ground states.

\section{Non-abelian adiabatic mixing with local detunings}\label{non-abelian}
Aside from the conventional adiabatic path mentioned above, a new scheme known as non-abelian adiabatic mixing was proposed in Refs. \cite{WYW20, WZ84}. In this section, we will briefly review the non-abelian adiabatic mixing and its application. During the process of non-abelian adiabatic mixing, there is a transition between ground states under the independent set Hamiltonian 
\begin{equation}
    H_{in}=V\sum_{(i,j)\in E} (1+\sigma^z_i)(1+\sigma^z_j)
\end{equation} 
if we apply single qubit rotations to each atom starting from the initial state $\ket{g_1 g_2 \dotsi g_N}$, where the $\sigma^z_i$ refers to the Pauli $Z$ matrix on the $i^{th}$ atom. These rotations generate a non-abelian gauge matrix $A$, and this matrix acts as an emergent Hamiltonian. The time evolution of the initial state is then given by
\begin{equation}
    \ket{\Psi(t)}=\mathcal{T} \bigg[ \exp (-i\int^t_0 A (t')dt') \ket{\Psi_{in}}   \bigg],
\end{equation}
where $\mathcal{T}$ is the time ordering operator. In Refs. \cite{YWW21}, the authors figured out that MIS is the ground state of $A(\theta=\pi)$ where
\begin{align}
    &A_{\mu,\nu}(\theta,\phi)  = i\bra{E_\mu}\partial\ket{E_\nu}\\
    &= \begin{cases}
        \frac{\sin\theta}{2}\frac{d\varphi}{dt}+\frac{i}{2}\textrm{sgn}(\mu-\nu)\frac{d\theta}{dt}  & \mu\neq\nu \\
        -\bigg(n_+ \sin^2 \frac{\theta}{2}+(n-n_+)\cos^2\frac{\theta}{2}  \bigg) \frac{d\varphi}{dt}   & \mu=\nu
    \end{cases}
\end{align}
under the condition $d\theta/dt \ll d\varphi_g/dt$. The symbols $\theta$ and $\varphi$ refer to conventional single qubit rotation angles on the Bloch sphere. Using these equations, the authors in Ref. \cite{ZGY+23} proved that this non-abelian adiabatic mixing and the Rydberg Hamiltonian are mathematically equivalent. The waveforms $\Omega(t),~\Delta(t)$, and $\phi(t)$ correspond to the non-abelian adiabatic mixing is as follows,
\begin{align}
    & \Omega(t)=\sqrt{w_\varphi^2 \sin^2(w_\theta t)+w_\theta^2}, \\
    & \Delta(t) = w_{\varphi} \cos(w_\theta t),\quad \tn{and}\\
    & \phi(t) = -\arctan \frac{w_\theta}{w_\varphi\sin(w_\theta t)}+\pi,
\end{align}
where the symbol $w_\varphi=d\varphi/dt$ and $w_\theta=d\theta/dt$. Given that Pauli matrices on different qubits are commute, we can directly apply our local detunings scheme to this non-abelian adiabatic mixing. Therefore, the complete adiabatic path of our strategy with the perspective of non-abelian adiabatic mixing must be local for all waveforms as follows:
\begin{align}
    & \Omega_i^g(t)=\sqrt{(w_\varphi^i)^2 \sin^2(w_\theta t)+w_\theta^2}, \\
    & \phi_i^g(t) = -\arctan \frac{w_\theta}{w_\varphi^i\sin(w_\theta t)}+\pi,
\end{align}
where $w_\varphi^i=w_\varphi-\epsilon(\mathbb{E}(\mathbf{s})-s_i))$ which is similarly defined to Eq. (\ref{weighted_hamiltonian}). The value of $w_\theta$ is fixed as $w_\theta T=\pi$ to make sure the constraint $A(\theta(T)=\pi)$ is satisfied.

\begin{figure}[h]
	\centering
	\includegraphics[width=0.7\linewidth]{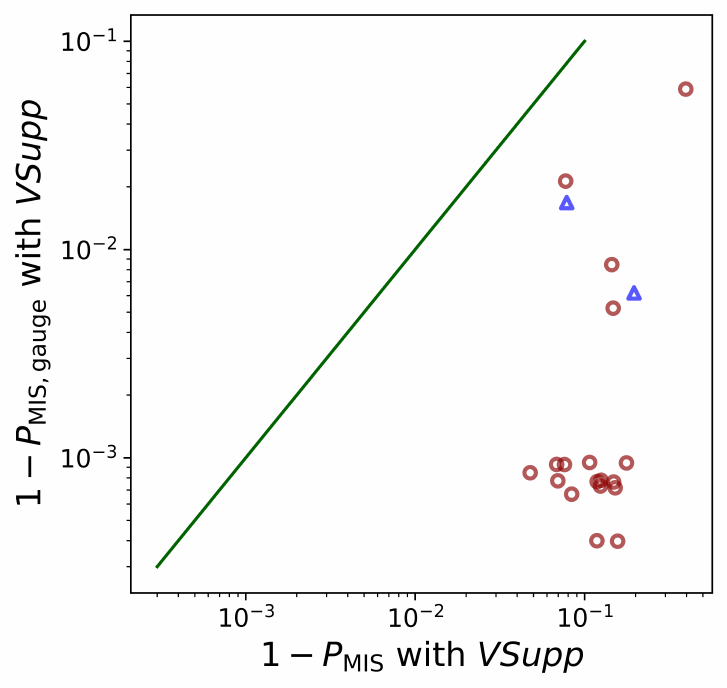}	
	\caption{Comparing the $1-P_{\mathrm{MIS}}$ `with \textit{VSupp}' between the waveforms $\{\Omega_i^g(t),\Delta^g_i(t),\phi_i^g(t)\}$ obtained from the non-abelian adiabatic mixing and the waveforms $\{\Omega(t), \Delta_{i,\mathrm{start(end)}}(t)\}$ on twenty $5\times 5$ checkerboard graphs with defects. Two blue triangles refer to their $\epsilon$ being negative when we optimize $\{w_\varphi, \epsilon\}$.}
	\label{fig:gauge_p}
\end{figure}

With these new waveforms, we optimize $\{w_\varphi,~\epsilon\}$ to obtain maximum $P_{\mathrm{MIS,gauge}}$ on $5\times5$ checkerboard graphs with defects. In addition, we compare these results to the $P_{\mathrm{MIS}}$ obtained using waveforms
\begin{align}
    & \Omega(t)=\sin(\pi t /T), \\
    & \Delta_{i,\mathrm{start(end)}}(t)=\Delta_{\mathrm{start(end)}}+\epsilon(\mathbb{E}(s)-s_i),\quad\tn{and} \\
    & \phi(t)=0.
\end{align}
The reason why we chose a sinusoidal form for $\Omega$ is that $\Omega_i^g(t)$'s smoothness may affect adiabaticity \cite{AK21}. More details on this point are in Appendix A and B. A comparison between the two waveforms is shown in Figure \ref{fig:gauge_p}. In addition, the classical post-processing was not used in this case to compare the true performance of these waveforms. The symbol $P_{\mathrm{MIS,gauge}}$ refers the probability of finding MIS using $\{\Omega_i^g(t),\Delta^g_i(t),\phi_i^g(t)\}$.

After $1~\mu s$ of time evolution, the values of $1-P_{\mathrm{MIS,gauge}}$ are significantly smaller than $1-P_{\mathrm{MIS}}$. Notably, many values of $1-P_{\mathrm{MIS,gauge}}$ are 100 times smaller than those of $1-P_{\mathrm{MIS}}$, which is an impressive result. Even in cases where $\epsilon$ are negative, which indicates that our strategy has failed, the values of $1-P_{\mathrm{MIS,gauge}}$ were still reduced. These two cases are marked with blue triangles in Figure \ref{fig:gauge_p}. This suggests that using the non-abelian adiabatic mixing is the way to extract the power of local detunings truly.

\section{Discussion}
\label{sec:con}
Finding maximum independent sets using Rydberg atom arrays has emerged as a potential candidate for achieving quantum advantage, although this advantage has yet to be fully demonstrated. In this paper, we proposed a new approximation strategy for finding unit-disk maximum independent sets on Rydberg atom arrays. This strategy is based on the assumption that a vertex with more connections to other vertices is less likely to be part of the maximum independent sets. By quantifying connectivity between vertices using vertex support and incorporating this into the Rydberg Hamiltonian, our approach successfully reduces the error rate by a factor of three for checkerboard graphs with defects, compared to the error rate of conventional QAA without local detunings. For random graphs with a density of 3.0, the error rate is successfully decreased compared to that obtained using the conventional QAA, regardless of both the degree of adiabaticity and the probability of obtaining ground states. Furthermore, our strategy can be extended to solve arbitrary connected maximum independent set problems on Rydberg atom arrays by various encoding strategy \cite{KKH+22, NLW+23, GMOS23}, as our assumption regarding connectivity does not rely on the property of the unit-disk graph.

Beyond its role in finding MIS on Rydberg atom arrays, the concept of vertex support can also aid in preparing specific quantum ground states when no classical post-processing is involved, which has potential applications including quantum error correction~\cite{SDM+22} and quantum communication~\cite{HBW+19}. For instance, applying this process to create a cat state on a $4\times 4$ square lattice increases the minimum energy gap during the evolution, thereby enhancing the fidelity between the prepared ground state and the desired 2D cat state from 0.68 to 0.90. This method can be applied to other lattices such as kagome lattice and honeycomb lattice. 

Additionally, we have applied our strategy to the non-abelian adiabatic mixing and compared the result to the conventional waveforms with local detunings. We have found that the values of $1-P_{\mathrm{MIS}}$ are much more reduced by the gauge field approach. This result is impressive given that we only need two parameters to control all $\{\Omega_i(t),\Delta_i(t),\phi_i(t)\}$. Even though experimentally realizing a set of local waveforms $\{\Omega_i(t),\Delta_i(t),\phi_i(t)\}$ is still far-reaching, this success indicates that there is plenty of room to improve the performance of our scheme, given that we only use constant $w_\varphi$ and $w_\theta$.

Our strategy can also be adapted for QAOA by encoding an amount of potential for vertex inclusion directly into the cost function \cite{EKC+22}. In fact, QAOA may be more suitable for our strategy than QAA because of its diabatic approach to preparing ground states \cite{ZWC+20}. Even though the MWIS cost function generated by our strategy may increase the population of the first excited state, diabatic evolution can actually result in a greater ground state population after crossing the minimum energy gap~\cite{SRN+12,CFL+14}. Our numerical calculations indicate a significant increase in $P_{\tn{MIS}}$ for complex graphs, supporting this argument. Since our strategy is based on the assumption, finding new weights that work better in representing connectivity between vertices will be our future research. Furthermore, our concept of utilizing vertex support for preparing quantum many-body ground states or obtaining ground state energy is versatile and can be adapted to various platforms and problems in a similar manner. In addition, exploring other applications of our approach is possible as our assumption is not limited to the maximum independent set problem alone.

\section*{Appendix A: Maximize $P_{\textrm{MIS}}$ without classical post-processing}
In the main paper, we maximize the total Rydberg density $\langle n\rangle = \sum_{i=1}^N \langle n_i\rangle$ to figure out maximum independent sets of graphs where $\langle n_i \rangle$ indicates the Rydberg density at $i^{th}$ site. Since maximizing $\langle n\rangle$ can lead the system to several Rydberg blockade violations, we perform the classical post-processing after the evolution. To demonstrate that our strategy facilitates the probability of getting low-level energy states, we maximize the $P_{\textrm{MIS}}$ without the classical post-processing in this section. This is impractical in the actual case when we do not know the maximum independent sets. However, it is a good indication of the ability of our strategy.

We present the $P_{\textrm{MIS}}$ of $5\times 5$ checkerboard graphs with defects, the same graphs exploited in the main paper. Other conditions, such as an evolving time, waveforms $\Omega(t),~\Delta(t)$, and initial conditions, are all the same. 

\begin{figure*}[ht]
	\centering	
	\begin{subfigure}{0.33\linewidth}
		\centering
		\includegraphics[width=\textwidth]{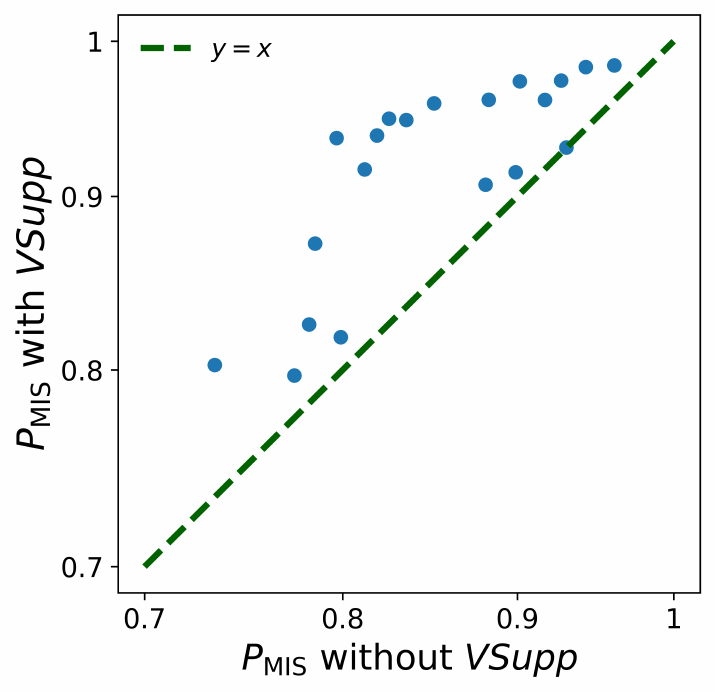}
		\caption{}
		\label{fig:without_processing}
	\end{subfigure}
	\hfil
	\begin{subfigure}{0.33\linewidth}
		\centering
		\includegraphics[width=\textwidth]{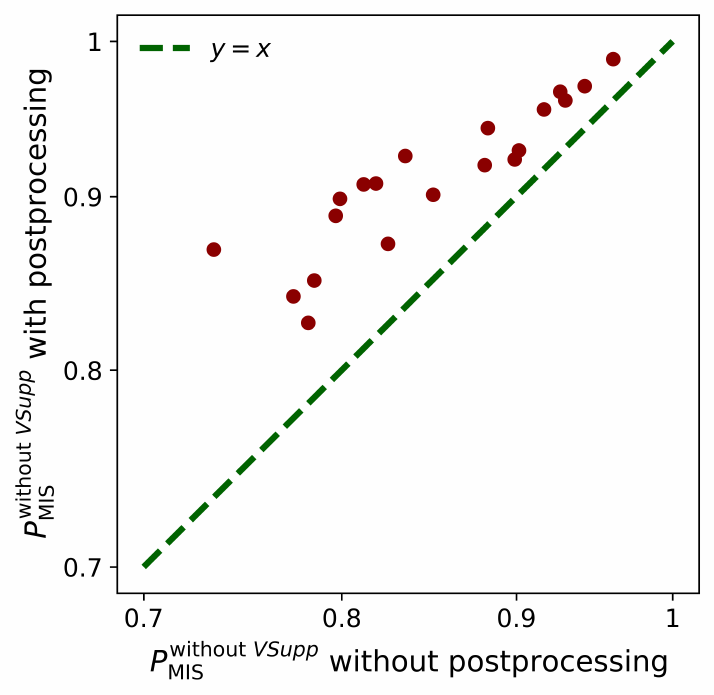} 
		\caption{}
		\label{fig:without_versus_processing}
	\end{subfigure}
	\hfil
	\begin{subfigure}{0.33\linewidth}
		\centering
		\includegraphics[width=\textwidth]{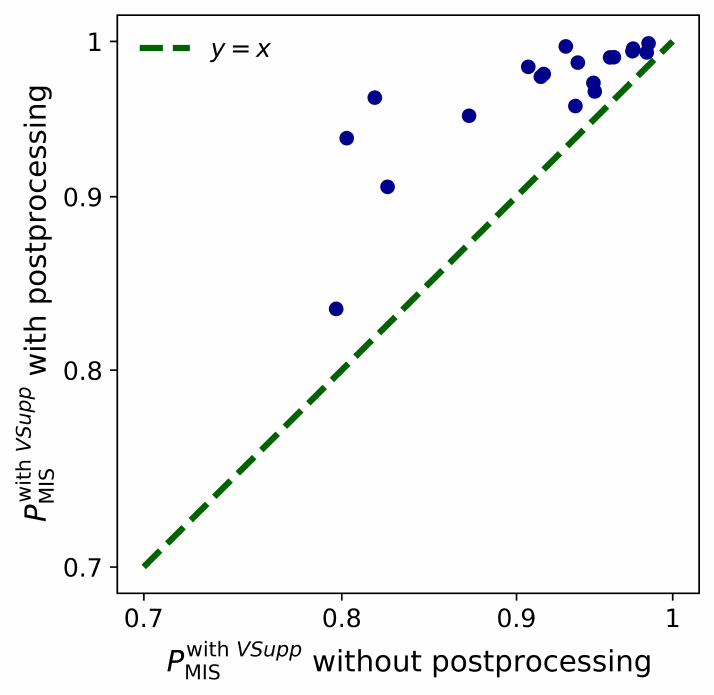}
		\caption{}
		\label{fig:with_versus_processing}
	\end{subfigure}
	\caption{Twenty checkerboard graphs with defects, $5\times5$, are used. a) Optimized $P_{\textrm{MIS}}$ without classical post-processing is described. The $x$-axis and the $y$-axis indicate the  $P_{\textrm{MIS}}$ `without \textit{VSupp}' and `with \textit{VSupp}', respectively. b) For the case of `without \textit{VSupp},' the $P_{\textrm{MIS}}$ without the classical post-processing is plotted against the  $P_{\textrm{MIS}}$ with post-processing. The $P_{\textrm{MIS}}^{\mathrm{without}~VSupp}$ refers the $P_{\textrm{MIS}}$ of `without \textit{VSupp}'. c) For the case of `with \textit{VSupp}', the  $P_{\textrm{MIS}}$ without classical post-processing is plotted against the $P_{\textrm{MIS}}$ with post-processing. The $P_{\textrm{MIS}}^{\mathrm{with}~ \textit{VSupp}}$ refers to the $P_{\textrm{MIS}}$ of `with \textit{VSupp}'.}
	\label{fig:with_versus_without}
\end{figure*}

The results of optimizing $P_{\textrm{MIS}}$ without the classical post-processing are described in Figure \ref{fig:with_versus_without}. In Figure \ref{fig:without_processing}, we can figure out that our strategy actually raises the $P_{\textrm{MIS}}$. By comparing the $P_{\textrm{MIS}}$ of 'without classical post-processing' and that of 'with classical post-processing' in Figure \ref{fig:without_versus_processing} and \ref{fig:with_versus_processing}, we can find the tendency that if the $P_{\textrm{MIS}}$ of 'without classical post-processing' is large, the $P_{\textrm{MIS}}$ of 'with classical post-processing' is also large.

\section*{Appendix B: Effects of smoothening $\Omega(t)$}
\begin{figure}[ht]
	\centering	
	\includegraphics[width=0.4\textwidth]{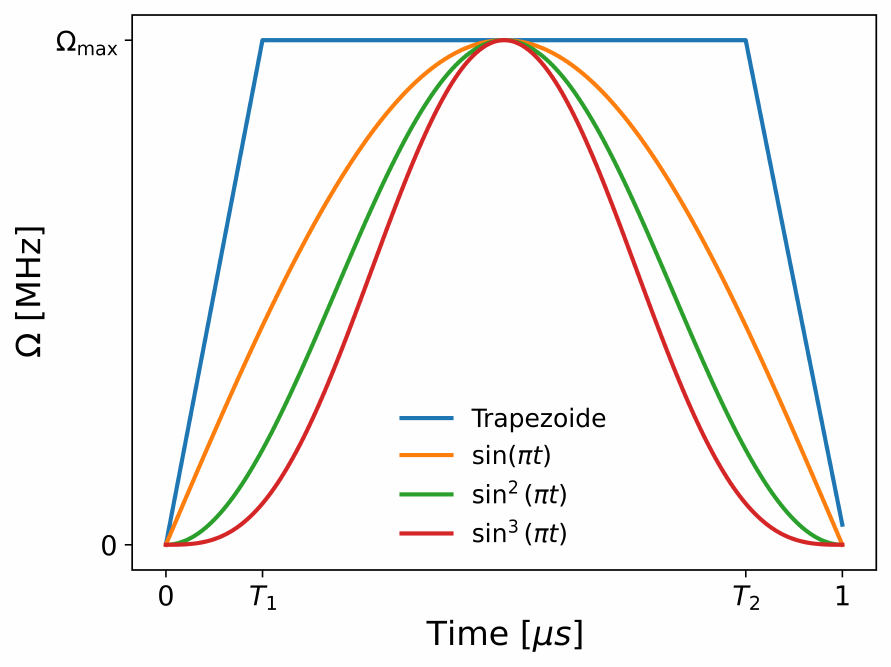}
	\caption{Description of used four $\Omega(t)$; a trapezoidal shape, $\sin(\pi t)$, $\sin^2(\pi t)$, and $\sin^3(\pi t)$.}
	\label{fig:four_omega}
\end{figure}

Due to the simplicity, we take a trapezoidal waveform $\Omega(t)$ in the main paper. This waveform has the advantage of a long evolution time. However, as a trade-off, there are abrupt changes at the beginning and the end of the evolution. This sudden variant may harm the adiabatic process because many adiabatic criteria basically require a slow change of the Hamiltonian \cite{AL18, A09}. According to the Ref. \cite{AK21}, if we explicitly put a non-adiabatic term in the non-degenerated Hamiltonian as a perturbation term and derive the first-order eigenstates of that Hamiltonian, we can get following time-dependent eigenstates $\{|n^{(1)}(t)\rangle\}$ about zeroth order eigenstates $\{|n^{(0)}(t)\rangle\}$ and eigenvalue $\{E^{(0)}_n\}$,
\begin{equation}
	|n^{(1)}(\tau)\rangle = |n^{(0)}(\tau)\rangle + \epsilon\sum_{k\neq n} M^{(0)}_{k,n}(\tau)|k^{(0)}(\tau)\rangle + O(\epsilon^2),
\end{equation}
where $\epsilon=\frac{1}{T}$, $\tau\in[0,~1]$, and
\begin{equation}
	 M^{(0)}_{k,n} \equiv -i\frac{\langle k^{(0)}(\tau)|\frac{\partial}{\partial_\tau}|m^{(0)}(\tau)\rangle}{E^{(0)}_n(\tau)-E^{(0)}_k(\tau)}.
\end{equation}
Harnessing these equations, we can demonstrate that if the Hamiltonian's changing rates are zero and the Hamiltonian is continuous at the beginning and the end, the error at the start and the end of the evolution is the second order of the inverse of total evolution time $\epsilon$. In this perspective, we analyze the effects of smoothening $\Omega(t)$ by taking four shapes of $\Omega(t)$: a trapezoidal shape, $\sin(\pi t)$, $\sin^2(\pi t)$, and $\sin^3(\pi t)$. The maximum values of $\Omega(t)$ are identical to $\Omega_{\textrm{max}}=4\pi\times 2.0~\textrm{MHz}$. We let the waveform $\Delta_i (t)$ remain the same because they already don't move at the beginning and the end. The four $\Omega(t)$ are described in Figure \ref{fig:four_omega}.

\begin{figure*}[ht]
	\centering	
	\begin{subfigure}{0.33\linewidth}
		\centering
		\includegraphics[width=\textwidth]{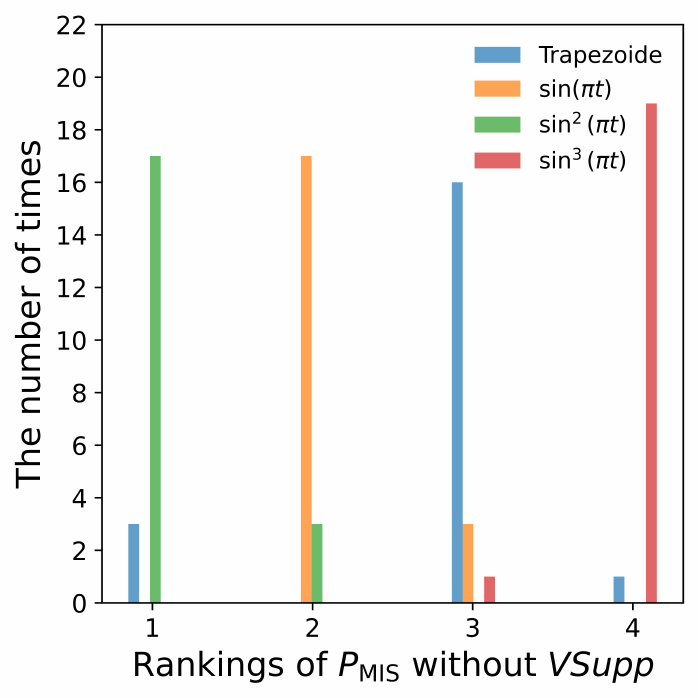}
		\caption{}
		\label{fig:ranking_without}
	\end{subfigure}
	\hfil
	\begin{subfigure}{0.33\linewidth}
		\centering
		\includegraphics[width=\textwidth]{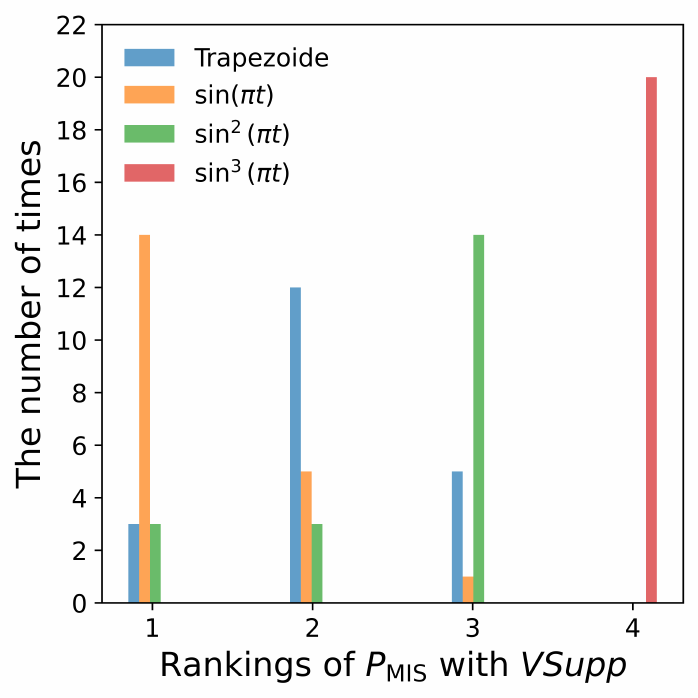} 
		\caption{}
		\label{fig:ranking_with}
	\end{subfigure}
	\hfil
	\begin{subfigure}{0.33\linewidth}
		\centering
		\includegraphics[width=\textwidth]{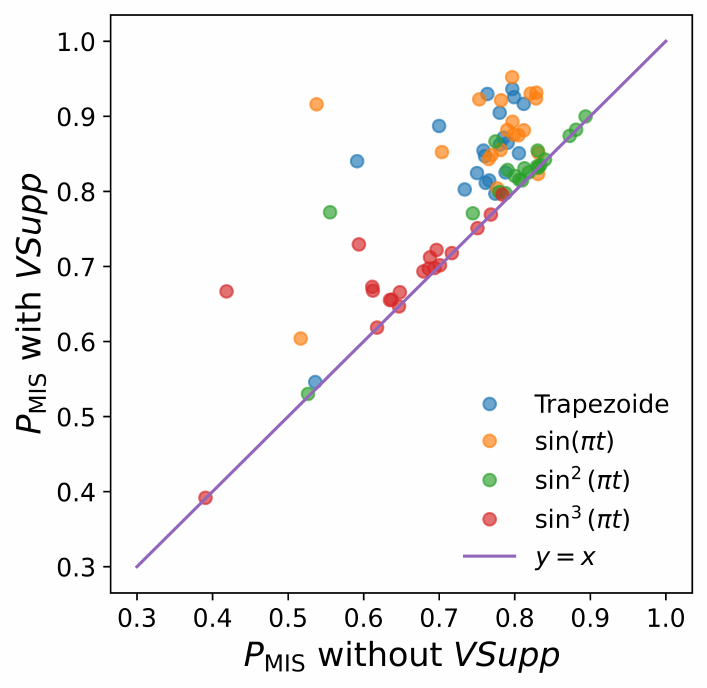}
		\caption{}
		\label{fig:compare_two}
	\end{subfigure}
	\caption{The twenty graphs, which are $5\times 5$ checkerboard graphs with defects with ground degeneracy 1, are used for (a), (b), and (c). a) After ranking in the order of the highest $P_{\textrm{MIS}}$ without classical post-processing for the `without \textit{VSupp}', the number of times each rank is plotted. b) After ranking in the order of the highest $P_{\textrm{MIS}}$ without classical post-processing for the `with \textit{VSupp}', the number of times each rank is plotted. c) Without classical post-processing, the $P_{\textrm{MIS}}$ `without \textit{VSupp}' is described against the $P_{\textrm{MIS}}$ `with \textit{VSupp}'.}
	\label{fig:effect_of_smoothe}
\end{figure*}

We utilize these waveforms in the quantum evolution of 20 checkerboard graphs, $5\times5$ with defects, with ground degeneracy 1 to maximize $P_{\textrm{MIS}}$ without the classical post-processing. After that, we compare two strategies, `without \textit{VSupp}' and `with \textit{VSupp}' in Figure \ref{fig:effect_of_smoothe}. Figure \ref{fig:ranking_without} and \ref{fig:ranking_with} describe the number of times the ranking of how large the $P_{\textrm{MIS}}$ of each waveform is. In Figure \ref{fig:ranking_without}, the waveform $\sin^2(\pi t)$ gets the most first place in the $P_{\textrm{MIS}}$ `without \textit{VSupp}', supporting the above analysis. Afterward, the waveform $\sin(\pi t)$ follows $\sin^2(\pi t)$. The waveform $\sin^3(\pi t)$ comes in last place the most, possibly because of the short evolution time due to the too many smoothenes. Interestingly, the results are reversed in Figure \ref{fig:ranking_with} and \ref{fig:compare_two}. The $P_{\textrm{MIS}}$ of waveforms $\sin(\pi t)$ and trapezoide are generally larger than that of $\sin^2(\pi t)$. Moreover, the $P_{\textrm{MIS}}$ of the waveform $\sin^2(\pi t)$ are less raised by our strategy than the $P_{\textrm{MIS}}$ of the waveform $\sin(\pi t)$ and trapezoid, as plotted in Figure \ref{fig:compare_two}. These results may imply that our strategy compensates for abrupt variations at the beginning and the end of the evolution.

\section*{Acknowledgements}
This work was supported by the National Research Foundation of Korea (NRF) through a grant funded by the Ministry of Science and ICT (NRF-2022M3H3A1098237, RS-2023-00211817), the Institute for Information \& Communications Technology Promotion (IITP) grant funded by the Korean government (MSIP) (No. 2019-000003; Research and Development of Core Technologies for Programming, Running, Implementing, and Validating of Fault-Tolerant Quantum Computing Systems), and Korea Institute of Science and Technology Information (P24021). H.E.K. acknowledges support by Creation of the Quantum Information Science R\&D Ecosystem through the National Research Foundation of Korea funded by the Ministry of Science and ICT (NRF-2023R1A2C1005588). 

\subsection*{Author Ccontributions}
H.Y. conceived and suggested the main idea. H.Y., H.E.K., and K.J. performed and analyzed the research, and wrote the manuscript.

\subsection*{Conflict of Interest}
The authors declare no conflict of interest.


\subsection*{Data Availability Statement}
The datasets and the code used during the current study available from the corresponding author on reasonable requests.

\subsection*{Keywords}
Quantum simulation, Rydberg atom arrays, Maximum independent set, Quantum adiabatic algorithm, Connectivity of graph


%

\end{document}